\documentclass[12pt]{article}
\usepackage{fullpage,authblk,cite}
\usepackage{amsmath,amsthm,graphicx,amssymb,verbatim,listings}
\usepackage{wasysym,color,enumitem,mathcomp}
\usepackage[T1]{fontenc}     % needed for the guillemets
\usepackage{lmodern, tipa}   % needed to fix suboptimal font rendering
\usepackage[ pdftex, plainpages = false, pdfpagelabels,
                 pdfpagelayout = useoutlines,
                 bookmarks,
                 bookmarksopen = true,
                 bookmarksnumbered = true,
                 breaklinks = true,
                 linktocpage=all,
                 pagebackref=false,
                 colorlinks = true,
                 linkcolor = Blue,
                 urlcolor  = blue,
                 citecolor = BrickRed,
                 anchorcolor = green,
                 hyperindex = true,
                 hyperfigures
                 ]{hyperref}
\usepackage[usenames, dvipsnames]{xcolor}
\usepackage{xifthen} % provides \isempty test

\newcommand{\tr}{\hbox{tr}}

\newcommand{\ket}[1]{{\ensuremath{\left| #1 \right\rangle}}}

\newcommand{\ketbra}[2]{{\ensuremath{\left| #1 \middle\rangle \middle\langle #2
      \right|}}}

\newcommand{\arxiv}[2][]{\ifthenelse{\isempty{#1}}{\href{http://arxiv.org/abs/#2}{{\tt arXiv:\allowbreak{}#2}}} {\href{http://arxiv.org/abs/#2}{{\tt arXiv:\allowbreak{}#2 [#1]}}}}

\newcommand{\booktitle}{\textsl}
\newcommand{\hrefdoi}[2]{\href{https://dx.doi.org/#1}{#2}}

\newcommand{\bbC}{\mathbb{C}}

\begin{document}
\title{Contradictions or Curiosities? On Kent's Critique of the Masanes--Galley--M\"uller Derivation of the Quantum Measurement Postulates}
\author[$\dag$]{Blake C.\ Stacey}
\affil[$\dag$]{Physics Department, University of
    Massachusetts Boston\protect\\ 100 Morrissey Boulevard, Boston MA 02125, USA}

\date{\small\today}

\maketitle

\begin{abstract}
  Adrian Kent has recently criticized Masanes, Galley and M\"uller's
  work on postulates for quantum mechanics. MGM claim to find two
  contradictions in Kent's criticism. I argue that neither is a true
  contradiction unless some other premise is added.
\end{abstract}

The story begins with ``The measurement postulates of quantum
mechanics are operationally redundant'' by Masanes, Galley and
M\"uller~\cite{Masanes:2019}. Then (skipping a bit) came ``The
measurement postulates of quantum mechanics are not redundant'' by
Kent~\cite{Kent:2023}. In turn, MGM wrote a
reply~\cite{Masanes:2023}. MGM assume that each physical system is
associated with a Hilbert space $\bbC^d$ and the ``pure states of the
system'' are the rays of that Hilbert space (i.e., vectors up to
overall multiplicative factors). They also assume that the reversible
transformations of the pure states are the unitary transformations
of~$\bbC^d$ and that state spaces compose by the tensor
product. Moreover, they introduce a ``state estimation'' postulate
which says, roughly, that the probability of any measurement outcome
can be calculated knowing the probabilities of some finite set of
outcomes. From these assumptions, they aim to deduce that measurement
outcomes are represented by linear operators, that probabilities for
measurement outcomes are calculated via the Born rule and that
post-measurement state update is specified by a completely positive
map associated with the measurement outcome. Kent argues that the
assumptions are not strong enough to reach the desired conclusion.

Kent introduces a ``state readout'' measurement that reports the
partial trace of a state defined on a multipartite Hilbert space. If
$\ket{\psi}$ is a unit vector in $\bbC^{d_1} \otimes \cdots \otimes
\bbC^{d_n}$, then the output of a state readout device is the partial
trace of $\ketbra{\psi}{\psi}$ over all the factors but the first, and
so it is a trace-1 positive semidefinite operator on the Hilbert space
$\bbC^{d_1}$. Which basis this matrix is reported in might depend upon
the construction of the measurement device or a set of input
parameters; the details of that aspect are not significant. If the
vector $\ket{\psi}$ is a tensor product, then the output will be a
rank-1 projector. Otherwise, it will be some matrix $\rho$ for which
$\tr \rho^2 < 1$.

Kent argues that this nonquantum measurement is consistent with all of
MGM's stated premises, and therefore those premises do not suffice to
imply the desired conclusions.

Using $\mathsf{A},\mathsf{B}$ to denote systems and $\mathcal{A},
\mathcal{B}$ to denote their associated Hilbert spaces, MGM write,
\begin{quote}
  The outcome of $\mathsf{M}_{\mathsf{sr}}$ on any ray of the Hilbert
  space $\mathcal{A}$ is a unit-rank density matrix, and the same is
  true for any proper mixture of rays from $\mathcal{A}$. However,
  when systems $\mathsf{A}$ and $\mathsf{B}$ are entangled the outcome
  of [the state readout measurement] is a density matrix with rank
  higher than one. This implies that the pure states of system
  $\mathsf{A}$ must include elements that are not rays of
  $\mathcal{A}$. [\ldots] More specifically, the set of pure states of
  $\mathsf{A}$ is the set of density matrices acting on $\mathcal{A}$,
  since each of these produces an outcome probability that cannot be
  written as the mixture of other outcome probabilities.
\end{quote}
In short, they argue, the introduction of Kent's ``state readout''
device breaks the assumption that pure states are rays in
$\bbC^d$. But this is incorrect. Specifying a device
$\mathsf{M}_{\mathsf{sr}}$ does not give a function from matrices to
matrices, but from rays of a Hilbert space to matrices on a smaller
Hilbert space. There is no mathematical meaning to the question of
what $\mathsf{M}_{\mathsf{sr}}$ yields given a density matrix on
$\mathcal{A}$, because that is not how $\mathsf{M}_{\mathsf{sr}}$ is
\emph{defined.} Given a specification of a state readout device and a
unit-trace positive semidefinite matrix, we cannot say what the output
of that device will be.  If the matrix is not already a rank-1
projector, then the only way to get an answer at all is to provide
more information. We can either pick a preferred decomposition into a
convex combination of projectors, or we can construct a projector on a
bigger Hilbert space of which the given matrix is a partial trace. In
other words, the entities that in quantum theory we call mixed-state
density matrices provide the \emph{outputs} of
$\mathsf{M}_{\mathsf{sr}}$, not its inputs. The existence of
$\mathsf{M}_{\mathsf{sr}}$ is consistent with their assumption about
pure states after all.

The idea that a pure state is a state that cannot be written as a
convex combination of other states is common in operational approaches
to quantum theory (see, e.g., \cite{Barnum:2023, Weiss:2023} and
references therein). But if one assumes, as MGM do, that pure states
are rays in $\bbC^d$, then an operational definition of ``pure state''
is beside the point: The set is already established. All of Kent's
hypothetical post-quantum devices are defined by their action on pure
states, which by assumption are rays in a Hilbert space, so trying to
deduce what the pure states are from the action of the post-quantum
devices is a meaningless exercise in contortionism.

MGM want Kent's state-readout device to contradict their postulates,
but what they present is a peculiarity rather than a
contradiction. And it is no surprise that a theory including a
state-readout machine would be peculiar, since by construction it is
not quantum theory.

The second claimed contradiction involves Kent's idea of a
``stochastic positive operator readout device'' (which Kent
abbreviates \emph{SPOD} and which MGM denote with
$\mathsf{M}_{\mathsf{spo}}$). This device works like a POVM, in that
it is specified by a resolution of the identity into positive
semidefinite operators and its outcome probabilities are found by the
Born rule. However, it does not alter the state that describes the
system. A theory that includes such devices will naturally exhibit
nonquantum behavior when sequences of multiple measurements are concerned.

MGM take up this theme, but in a way that brings in a new premise:
\begin{quote}
  However, within our operational framework, a measurement is any
  valid process which produces a classical outcome. In particular, we
  can consider the joint outcome $(i, j)$ of first, apply measurement
  $\mathsf{M}_{\mathsf{spo}}$ obtaining outcome $i$, and second, apply
  an ordinary POVM measurement obtaining outcome $j$.
\end{quote}
If a measurement is ``any valid process which produces \emph{a}
classical outcome'' (emphasis added), then it is not clear that the
idea of a ``joint outcome'' even makes sense. Once we obtain the
outcome $i$, we have an outcome in hand; the measurement is
complete. We can \emph{assume} that a two-step measurement is not
radically different from a one-step measurement, but this is a
nontrivial assumption. This has bothered me before~\cite{Stacey:2022},
but the phrasing here underlines the problem. Kent explicitly states
that in his reading, ``MGM do not assume that a sequence of
measurements can be considered as a single measurement''. Perhaps such
an assumption was \emph{intended} all along; in their response to my
earlier comment, MGM write, ``Our derivation of the Born rule is
completely agnostic as to how the outcome comes about or how many
steps it takes''~\cite{Masanes:2022}. This is a fine assumption to
include, but it just wasn't there as written. And if one does wish to
rely upon a premise of this type, it needs to be spelled out. It's
very easy to go too far and assume a level of agnosticism that quantum
theory cannot sustain, whether in blunt ways (like eliminating
interference phenomena) or more subtle~\cite{Leifer:2013,
  Horsman:2017}. By analogy, we might say that the key assumption of
Gleason's theorem was that the probability associated to a ray is
``agnostic'' about which orthonormal set that ray is embedded
within~\cite{Gleason:1957}. Gleason was explicit about exactly how to
be agnostic; the same philosophy applies here.

MGM then pose their second claimed contradiction:
\begin{quote}
  In what follows we show that such a composite measurement has a
  probabilistic structure different than that of quantum theory.
\end{quote}
This, by itself, is neither problematic nor surprising. The stochastic
positive operator readout device $\mathsf{M}_{\mathsf{spo}}$ is a
nonquantum ingredient, and so using it in a recipe will yield a
nonquantum confection. Its purpose in Kent's paper is to construct a
situation where the state-update rule cannot be written as a
completely positive map independent of the initial state. MGM claim
that this ``invalidates'' Kent's refutation of their derivation, but
Kent already made the same mathematical conclusion. The only
contradiction is with the idea that a ``joint outcome'' can always be
treated like a single POVM element. In quantum theory, this is true,
but in a nonquantum theory it might be false. Nothing logically
forbids us from imagining that a laboratory protocol can be an
arbitrarily long sequence of POVMs and SPODs.

\medskip

MGM employ a premise that they designate ``the possibility of state
estimation''. This premise concerns probabilistic mixtures of pure
states, i.e., scenarios where pure state $\psi_r$ is present with
probability $p_r$ as the index $r$ ranges over some set. For MGM,
``the possibility of state estimation'' is the assumption that some
set of measurement outcomes exists with the property that knowing the
probabilities for these outcomes implies the probabilities for all
others. In other words, a finite set of probabilities (not necessarily
all pertaining to the same measurement) is all that we need to
calculate the statistics of any experiment for any mixture $\{(p_r,
\psi_r)\}$. This is true in quantum theory: Any set of outcomes that
correspond to operators which span the space of positive semidefinite
matrices on~$\bbC^d$ is informationally complete~\cite{DeBrota:2020,
  Slomczynski:2020, DeBrota:2021}. However, if we introduce a state
readout device, then we obtain a nonquantum theory in which this
method of specifying a mixture by a finite list of probabilities
cannot work. A finite list of probabilities is enough to specify a
density matrix, but as noted above, the output of a state readout
device depends upon a particular convex decomposition of that density
matrix into rank-1 projectors.

That said, having a device which reads off a state vector seems like
it ought to be an aid to tomography, rather than an impediment. Kent
observes that we could use the state readout device to identify the
rays $\{\psi_r\}$ and gather statistics to arrive at values for the
corresponding $\{p_r\}$. So, to express what we mean by ``the
possibility of state estimation'', we ought to make a mathematical
assumption that is less restrictive, or that is adapted to the
measurement set that we have available.

Kent also criticizes MGM's ``possibility of state estimation''
postulate for being expressed in terms of probabilistic mixtures when
by the earlier assumptions ``measurement outcome functions are
uniquely defined by their action on pure states''. For Kent, mixtures
``need not necessarily have any fundamental status'' like pure states
do. This, I suspect, is an interpretation-dependent intuition. If one
regards a quantum state for a system as an encoding of expectations
regarding that system, then pure states need not have a more
fundamental status. In such a view, a pure state is merely a maximally
focused catalogue of expectation, mathematically interesting for being
on the boundary of the full state space but not granted a different
ontological status from any state on the interior. Different choices
of interpretation will make different moves seem natural. Fully
committing to an epistemic or doxastic interpretation of quantum
states may facilitate taking steps to close off Kent's criticisms. For
example, it may motivate enforcing the condition that the
post-measurement state update map is always linear, by inheriting the
linearity of probability theory~\cite{Pienaar:2023}. That said, by the
same token it would also call into question the value of starting with
the assumption that the state space is furnished by the rays
of~$\bbC^d$ rather than by linear operators upon it.

\bigskip

This work was supported by National Science Foundation Grant PHY-2210495.


\begin{thebibliography}{999}
\bibitem{Masanes:2019} L.\ Masanes, T.\ D.\ Galley and M.\ P.\ M\"uller,
  ``\hrefdoi{10.1038/s41467-019-09348-x}{The measurement postulates of
  quantum mechanics are operationally redundant},'' \booktitle{Nature
  Communications} \textbf{10} (2019), 1361, \arxiv{1811.11060}.

\bibitem{Kent:2023} A.\ Kent, ``The measurement postulates of quantum
  mechanics are not redundant,'' \arxiv{2307.06191} (2023).

\bibitem{Masanes:2023} L.\ Masanes, T.\ D.\ Galley and
  M.\ P.\ M\"uller, ``Response to `The measurement postulates of
  quantum mechanics are not redundant','' \arxiv{2309.01650} (2023).

\bibitem{Barnum:2023} H.\ Barnum, C.\ Ududec and J.\ van de Wetering,
  ``Self-duality and Jordan structure of quantum theory follow from
  homogenity and pure transitivity,'' \arxiv{2306.00362} (2023).

\bibitem{Weiss:2023} M.\ B.\ Weiss, ``Depolarizing reference devices
  in Generalized Probabilistic Theories,'' \arxiv{2312.12790} (2023).
  
\bibitem{Stacey:2022} B.\ C.\ Stacey, ``Masanes--Galley--M\"uller and
  the state-update postulate,'' \arxiv{2211.03299} (2022).

\bibitem{Masanes:2022} T.\ D.\ Galley, L.\ Masanes and
  M.\ P.\ M\"uller, ``Reply to `Masanes--Galley--M\"uller and the
  State-Update Postulate','' \arxiv{2212.03629} (2022).
  
\bibitem{Leifer:2013} M.\ S.\ Leifer and R.\ W.\ Spekkens,
  ``\hrefdoi{10.1103/PhysRevA.88.052130}{Towards a formulation of
  quantum theory as a causally neutral theory of Bayesian
  inference},'' \booktitle{Physical Review A} \textbf{88} (2013),
  052130, \arxiv{1107.5849}.

\bibitem{Horsman:2017} D.\ Horsman, C.\ Heunen, M.\ F.\ Pusey,
  J.\ Barrett and R.\ W.\ Spekkens,
  ``\hrefdoi{10.1098/rspa.2017.0395}{Can a quantum state over time
    resemble a quantum state at a single time?},''
  \booktitle{Proceedings of the Royal Society A} \textbf{473} (2017),
  20170395, \arxiv{1607.03637}.

\bibitem{Gleason:1957} A.\ M.\ Gleason,
  ``\hrefdoi{10.1512/iumj.1957.6.56050}{Measures on the closed
  subspaces of a Hilbert space},'' \booktitle{Indiana University
  Mathematics Journal} \textbf{6} (1957), 885--93.

\bibitem{DeBrota:2020} J.\ B.\ DeBrota and B.\ C.\ Stacey,
  ``\hrefdoi{10.1103/PhysRevA.102.032221}{Discrete Wigner Functions
  from Informationally Complete Quantum Measurements},''
  \booktitle{Physical Review A} \textbf{102} (2020), 032221,
  \arxiv{1912.07554}.

\bibitem{Slomczynski:2020} W.\ S{\l}omczy\'nski and A.\ Szymusiak,
  ``\hrefdoi{10.22331/q-2020-09-30-338}{Morphophoric POVMs,
  generalised qplexes, and 2-designs},'' \booktitle{Quantum}
  \textbf{4} (2020), 338, \arxiv{1911.12456}.
  
\bibitem{DeBrota:2021} J.\ B.\ DeBrota, C.\ A.\ Fuchs and
  B.\ C.\ Stacey, ``\hrefdoi{10.1142/S0219749920400055}{The Varieties
    of Minimal Tomographically Complete Measurements},''
  \booktitle{International Journal of Quantum Information} \textbf{19}
  (2021), 2040005, \arxiv{1812.08762}.

\bibitem{Pienaar:2023} J.\ L.\ Pienaar, ``Quantum dynamics is linear
  because quantum states are epistemic,'' \arxiv{2302.13421} (2023).
    
  \begin{comment}
\bibitem{Davies:1970} E.\ B.\ Davies and J.\ T.\ Lewis,
  ``\hrefdoi{10.1007/BF01647093}{An operational approach to quantum
  probability},'' \booktitle{Communications in Mathematical Physics}
  \textbf{17} (1970), 239--60.

\bibitem{Fuchs:2014} C.\ A.\ Fuchs, M.\ Schlosshauer (foreword) and
  B.\ C.\ Stacey (editor). \booktitle{My Struggles with the Block
    Universe,} \arxiv{1405.2390} (2014).

\bibitem{Flatt:2017} K.\ Flatt, S.\ M.\ Barnett and S.\ Croke,
  ``\hrefdoi{10.1103/PhysRevA.96.062125}{Gleason--Busch theorem for
  sequential measurements},'' \booktitle{Physical Review A}
  \textbf{96} (2017), 062125.
  
\bibitem{Busch:2003} P.\ Busch,
  ``\hrefdoi{10.1103/PhysRevLett.91.120403}{Quantum States and
  Generalized Observables:\ A Simple Proof of Gleason's Theorem},''
  \booktitle{Physical Review Letters} \textbf{91} (2003), 120403,
  \arxiv{quant-ph/9909073}.
  
\bibitem{Caves:2004} C.\ M.\ Caves, C.\ A.\ Fuchs, K.\ K.\ Manne and
  J.\ M. Renes,
  ``\hrefdoi{10.1023/B:FOOP.0000019581.00318.a5}{Gleason-Type
    Derivations of the Quantum Probability Rule for Generalized
    Measurements},'' \booktitle{Foundations of Physics} \textbf{34}
  (2004), 193--209, \arxiv{quant-ph/0306179}.

\bibitem{Haag:1978} R.\ Haag and U.\ Bannier,
  ``\hrefdoi{10.1007/BF01609470}{Comment on Mielnik's Generalized (Non
  Linear) Quantum Mechanics},'' \booktitle{Communications in
  Mathematical Physics} \textbf{60} (1978), 1--6.

\bibitem{Jordan:1993} T.\ Jordan,
  ``\hrefdoi{10.1006/aphy.1993.1053}{Reconstructing a Nonlinear
  Dynamical Framework for Testing Quantum Mechanics},''
  \booktitle{Annals of Physics} \textbf{225} (1993), 83--113.
  \end{comment}
\end{thebibliography}
\end{document}